\documentclass[12pt]{article}
\usepackage{graphicx}

\def\be{\begin{equation}}
\def\ee{\end{equation}}
\def\bea{\begin{eqnarray}}
\def\eea{\end{eqnarray}}
\def\ben{\begin{eqnarray*}}
\def\een{\end{eqnarray*}}

\def\d{\partial}
\def\e{\varepsilon}
\def\<{\langle}
\def\>{\rangle}
\def\nn{\nonumber\\}

\def\v#1{\mbox{\boldmath $#1$}}

\def\delete#1{}

\newtheorem{definition}{Definition}

\begin{document}

\title{%On the Quantum Mechanics of Relativistic Objects. part II:
	Quantization of Nambu-Goto Action in Four Dimensions}

\author{Tsuguo MOGAMI
	\thanks{e-mail: mogami@brain.riken.go.jp}\\
	RIKEN, Hirosawa 2-1, Wako-shi,
	Saitama, 351-0198 Japan \\
}

%\date{April 24, 2010}		%ver. 0.99
%\date{May 17, 2010}		% ver. 1.00, arXiv
\date{Oct 29, 2010}		% ver. 1.10, NPB

\maketitle

\begin{abstract}
Recently, a new quantization method for gauge theories was proposed, in which no gauge fixing is required but the constraints are kept.
Here we successfully applied this formalism to Nambu-Goto action in any dimensions.
The result of our theoretical calculation have shown striking agreement with the spectrum of the light-quark mesons.
\end{abstract}

%\section{ Introduction }

Quantization of strings has long been a hard problem in particle physics, even though the Regge behavior of mesons clearly suggested some stringy object should exist.  The string theory was also of interest as the fundamental string and much effort was made on it in vain.

Recently, a new method of quantizing gauge theories was proposed\cite{nonfix}, in which no gauge fixing was required.
Only obeying the equations of motion and constraints had given consistent result.  No trick of gauge fixing and BRST formalism were required.

In the conventional view, quantization is replacing the Poisson bracket $\{x, p\} = 1$ by the commutator $[x, p] /i \hbar = 1$.  This statement was originally from Dirac.  Later, Dirac noticed that this statement fails sometimes in the canonical formalism of constrained systems\cite{Dir64}.

When the constraints and the commutation relations are inconsistent, i.e.\ in a second-class system or when anomaly exists, we have two options.
The first is sustaining the constraint equations, but the commutation relations will be compromised, at least, partially.
The second is sustaining the commutation relations, but a part of the constraint equations will be given up.
There is no {\it a priori} answer to which option to take, and both of the options are seen in the literature, e.g.\ Dirac took an option of sustaining the constraints and invented Dirac brackets for this purpose.  The old covariant quantization (OCQ) of the string is an example of sustaining the commutation relations, in which only half of the constraint is taken, i.e.\ $L_m|*\> = 0$ only for non-negative $m$.

Therefore the idea of sustaining the constraints for strings too does worth pursuing at least to the end.
Then, we will fully sustain the constraints $L_m|*\> = 0$ for all positive and negative $m$.  Then the commutation relations should be partly given up.

Where can we change about the commutation relations?  We have 2 degrees of freedom that will be fixed by the constraints for each point $\sigma$ on a string.
Let $p_\mu$ be the total momentum of the string.
The string variables $X^\mu(\sigma)$ forms a Lorentz 4-vector, however, the component of $X^\mu(\sigma)$ that is proportional to $p_\mu$ will be always proportional to $p_\mu$ after Lorentz boost or rotation, which will be called $S$ component from here.
The other three directions are always orthogonal to $p_\mu$ and mix each other.
We have 2 degrees of freedom with $X^S(\sigma)$ and $\Pi^S(\sigma)$ for each point on the string, which is same as that of the constraints.
In contrast, the other three components have three times as many degrees of freedom, and they cannot be separated into any parts since they mix by Lorentz transformation.
Therefore we have to choose giving up the commutation relation for $S$ components while we spare the commutation relations for other components.

Applying this new method to Nambu-Goto string gave consistent result.  We had no negative norm states and no tachyons.
This theory can be used as either a fundamental string or a phenomenological string of hadrons, i.e.\ of the QCD.
Our theoretical calculation have shown striking agreement with the experimental spectrum of the light-quark mesons.

In section 1, quantization of the string action is performed, and
its representation is formed.  % One can read section 1 without reading section A if one only accepts eq.(\ref{comm}) and eqs.(\ref{constr}).
In section 2, the light-meson spectra are accounted.
In the appendix, it is explained why finding the minimal representation is enough as quantization, which is used in justifying equation eq.(\ref{comm}) and eqs.(\ref{constr}) in section 1.

Comments about the conventional methods will be put in the footnotes, since they are not important for the future readers.
We adopt a metric with signature $\{- + + + \cdots \}$, repeated indices are always summed, and $|*\>$ designates ``any physical state''.

\section{ Quantization of the String }

\subsection{ Classical formalism }

An open string may be described by $X^\mu (\sigma )$, where $\sigma $ is the spatial position along the string and ranges from $0$ to $\pi $.
Nambu-Goto action is \be
	L = \frac{1}{\pi} \int_0^\pi d\sigma  \sqrt{ \dot X^2 X'^2 - (\dot X\cdot X')^2}.
\ee
The conjugate momenta are defined by \[
	\Pi_\mu (\sigma ) = \frac{\delta L}{\delta \dot X^\mu (\sigma )}
	= { \dot X_\mu  X'^2 - X'_\mu (\dot X\cdot X')
		\over \pi \sqrt{ \dot X^2 X'^2 - (\dot X\cdot X')^2}
		}.
\]

Then the constraints for this system are \bea
	X'^\mu \Pi_\mu  = 0,			\\
	\pi^2 \Pi^\mu \Pi_\mu  + X'^\mu X'_\mu = 0
\eea
for all $\sigma $, and the Hamiltonian is $H = 0$.
Here we don't move to the canonical formalism.  $\Pi_\mu$ is used just as a convenient variables rather than $\dot X^\mu$ in expressing dynamics.

Let us consider the physical vibration modes in this classical theory before quantizing it.
Unless the total momentum is null $p^2 = 0$, the first vibrating solution for the Euler-Lagrange equations of motion is \be
	X^\mu(\sigma) = p^\mu \tau + \varepsilon^\mu \cos\sigma \cos\tau,
\ee
where $\varepsilon^\mu$ may be either \bea
	&& \varepsilon^L_\mu = (p_0, p_1, 0, 0)/p,	\\
	&& \varepsilon^+_\mu = (0, 0, 1, + i)/\sqrt{2},			\\
	&& \varepsilon^-_\mu = (0, 0, 1, - i)/\sqrt{2},
\eea
where $p \equiv \sqrt{-p^\mu  p_\mu }$ when the spatial component of the momentum is in $x$ direction.
If the polarization is in $S$ direction $\varepsilon^S_\mu = (p_0, p_1, 0, 0) /p $, this vibration is prohibited because the Lagrangian $L$ is zero for this direction, which means that this polarization belongs to the gauge freedom.
When $p^2 = 0$, there are also three directions of motions, which is different from the solution above.
Thus, in the classical theory, we have no limitation on space-time dimension $D$, and there is $D - 1$ directions of vibration for the first modes.
Our quantization method will give the same features, and therefore it is reasonable to think this method is the appropriate quantization of the classical theory.\footnote{
Quantized theory must coincide with the classical counterpart in the limit of $\hbar \rightarrow 0$.
Therefore the conventional string theory may not have classical counterpart because the number of vibration is different from the classical theory.
}

One may use the Polyakov action, which is also called Brink-Di Vecchia-Howe-Deser-Zumino action: \[
	S = -{1\over 2\pi} \int d\tau d\sigma  (-\gamma )^{1/2} \gamma^{ab} \d_a X^\mu  \d_b X_\mu .
\]
In fact, using it is simply equivalent because its equations of motion are simply equivalent to that of Nambu-Goto action, and our quantization is only finding a representation that conforms to the equations of motion (see Appendix A).

\subsection{ Construction of the State Space (Representation) }

The vibration may be decomposed into Fourier modes as \be
	X^\mu (\sigma ) = \sum_n \cos n\sigma \; X_n^\mu ,
\ee
and accordingly the canonical conjugate momenta may be decomposed as \be
	\Pi_n^\mu  = \int_0^\pi d\sigma  \cos n\sigma \; \Pi^\mu (\sigma ) .			\ee
%	\Pi^\mu (\sigma ) = {2\over \pi }\sum_n \cos n\sigma  \Pi_n^\mu .

Let us divide the variables $X_1^\mu , X_2^\mu , \cdots $ into \bea
	X_n^S =  X_{n,\mu } (p^0, p^1, 0, 0)^\mu /p ,			\\
	X_n^L =  X_{n,\mu } (p^1, p^0, 0, 0)^\mu /p,
\eea
where $p^\mu $ is the total momentum of the string, and $p \equiv \sqrt{-p^\mu  p_\mu }$, when the spatial component of the momentum is in $x$-axis.  The transverse components will be denoted by $X^+$ and $X^-$ (in four dimensions), and $X^T$ will designate either $X^+$ or $X^-$.
The definition of the $S$ component is Lorentz invariant because the $S$ component is defined as the scalar product of the variable $X_\mu$ and the total momentum $p^\mu$.
% Each of the four polarizations corresponding to these four components is orthogonal to the others.

We assume that the state space is constructed by multiplying $X^\mu $ and $\Pi^\mu $ onto the vacuum $|0\> $ as usual.  Let us assume that commutation relations are \bea
	[X_n^i, \Pi_n^j] = i \delta^{ij},
	\label{comm}
\eea
where $i$ and $j$ may take either $+$, or $-$ of the transverse modes or $L$, and note that we did not impose any condition on $[X_n^S, \Pi_n^{S}] $.  This will not cause any problem to Lorentz invariance of this model, because $X_n^S$ variables are Lorentz scalars and do not mix with $+, -,$ and $L$ polarizations.
Then we assume that all the constraints hold on any physical state $|*\> $: \bea
	(\pi^2 \Pi \cdot \Pi (\sigma ) + X'\cdot X'(\sigma ))|*\> = 0,			\nonumber\\
	\Pi \cdot X'(\sigma )|*\>  = 0
	\label{constr}
\eea
for all $\sigma $.
Eq.(\ref{comm}) and eqs.(\ref{constr}) are the basic assumptions of our quantization method, and it is inversely justified because it gives a consistent representation of the quantum states (see Appendix A).
Because we have as many constraints as to determine $X_n^S$'s and $\Pi_{n}^S$'s, we will shortly see that these variables may be determined and be written in terms of $X^L$, $X^T$, $\Pi^T$, and $\Pi^L$.
Then the state space may be constructed only with $X^L$, $X^T$, $\Pi^T$, and $\Pi^L$ operating onto the vacuum $|0\> $, which is equivalent to that the wavefunctions have only $X^L$, $X^T$, and $X^\mu_0$ in its arguments.
Further, the commutators that include $S$ modes will be determined by these relations.

The commutation relations (\ref{comm}) for $\alpha_n^T = \Pi_n^T - i n X_n^T/2$ and $\alpha_n^L = \Pi_n^L - i n X_n^L/2$ are \ben
	[\alpha_n^T, \alpha_n^{T\dagger }] &=& n ,			\\ \,
	[\alpha_n^L, \alpha_n^{L\dagger }] &=& n ,
\een
and the constraints are easier to handle using the combination \bea
	L_m &\equiv& \frac 1{2\pi} \int_0^\pi d\sigma  \{ \cos m\sigma (\pi^2 \Pi \cdot \Pi  + X'\cdot X') + 2\pi i \sin m\sigma \; X'\cdot \Pi  \}				\nn
	&=& {1\over 2} \sum_{n = -\infty }^\infty  \alpha_{m-n}\cdot \alpha_n.
\eea
Then eq.(\ref{constr}) may be rewritten as \be
	L_m|*\>  = 0
	\label{Lmvac}
\ee
for all\footnote{
In the conventional string theories, only the positive half of the conditions were possible to impose, which is incomplete to assure diffeomorhism invariance.
% Moreover the constraint equations are destroyed quantum mechanically by the conformal anomaly.
\par
This is the reason why we don't have the requirement $D = 26$.
In the old covariant quantization of strings, the reason why $D = 26$ is related to the negative norm states.  At first, the norm of the state $ ( L_{-2} + (3/2) L_{-1}^2 )|0\> $ is $(D - 26)/2$ and it should not be a negative norm.
In contrast, in our theory, both of the states $ L_{-1}|0\>$ and $ L_{-2} |0\> $ are equal to zero because of eq.(\ref{L-1vac}) and eq.(\ref{L-2vac}), which are direct consequences of sustaining all the constraints eq.(\ref{Lmvac}).  It is interesting that the number of the full constraints is exactly enough to erase the negative norm states.  % Therefore our method does not give any negative norm states and works in any spatial dimensions.
}
integer $m$.
Therefore we may use operator equations $ L_m = 0$ in the physical space.

The ground state is defined as \bea
	&& \alpha_n^T |0, p^\mu \> = 0 ,			\nn
	&& \alpha_n^L |0, p^\mu \> = 0 ,			\nn
	&& \Pi_0^\mu |0, p^\mu \> = p^\mu|0, p^\mu \>.
\eea
The total momentum $p^\mu$ in the ket vectors may be abbreviated from now on.  Here we have \bea
	&& \alpha_n^S |0, p^\mu \> = 0
\eea
for any positive $n$, where $\alpha_n^S = \Pi_n^S - i n X_n^S/2$.  This equation may be proven perturbatively up to any order in $1/p$ by repeatedly using $L_n = 0$\footnote{
This $L_n = 0$ for all $n$ means that $[L_m, L_n] = 0$ within the physical space.
Conventionally, this commutator should give an anomalous term called the ``central charge''.  The reason of not having this term is because eqs.(\ref{aSreduc}) are just writing $L_m = 0$ using $ S$ and the other components, and these equations for all negative and positive $m$ are solved in terms of $\alpha^S$'s.  The result is eq.(\ref{aS}).  Thus fulfilling $L_m = 0$ for all $m$ was possible.
\par
Further, there is no other choice than this determination of $ \alpha^S$'s, if we decide to keep the constraints and partially give up the commutation relations.  This is because of the counting of the numbers of the freedoms, which was discussed in the introduction.
}, i.e.\ \be
	\alpha_n^S = \frac{1}{p} \Big\{ L_n' - {1\over 2}\sum_{m= -\infty;\; m \ne 0, -n}^\infty \alpha_{-m}^S \alpha_{n+m}^S \Big\},
	\label{aSreduc}
\ee
to obtain \bea
	\alpha^S_n &=& {1\over p} L'_n - {1\over 2 p^3} \sum_{k \ne 0, n}  L'_{n-k} L'_k			\nn
	&& + {1\over 4 p^5} \Big(
		\sum_{k\ne 0,n;\; l \ne 0, n-k} L'_{n-k-l} L'_l L'_k
		+ \sum_{k\ne 0,n;\: l \ne 0, k} L'_{n-k} L'_{k-l} L'_l
	\Big) - \cdots
	\label{aS}
\eea
and using it, where $L_n' \equiv  {1\over 2}  \sum_i \sum_{n = -\infty, n\neq 0, m }^\infty  :\alpha_{m-n}^i\alpha_n^i: $, and Roman indices $i$ and $j$ designates $+, -,$ or $L$ when used as spatial indices.  For $m=0$, we have \bea
	L_0 &=& - \frac{p^2}{2} + L'_0 - {1\over 2 p^2} \sum_{k \ne 0} : L'_{-k} L'_k :			\nn
	&& + {1\over 4 p^5} \Big(
		\sum_{k\ne 0;\: l \ne 0, -k} L'_{n-k-l} L'_l L'_k
		+ \sum_{k\ne 0;\; l \ne 0, k} L'_{n-k} L'_{k-l} L'_l
	\Big) - \cdots
	\label{L0}
\eea
Here we suppose that $L'_k$'s are normal ordered in eq.(\ref{aS}) and eq.(\ref{L0}).
% Giving no ordering rule made numerically similar result.

% Indeed, this is not proven to be the unique representation for the equations of motion, we suppose that viable representation is unique.

\subsection{ The Spectrum }

In this subsection, let us list up the states and their energy.

The first state is obviously the vacuum \be
	|0\> .
\ee
It represents a spin-0 state and its invariant mass is $p^2 = - a$ because of the constraint $(L_0 -a)|0\> = 0$.  Here a negative constant $a$ was introduced to absorb the ordering ambiguity of $L_0$.

In the next level, the physical states are \be
	\alpha_{-1}^i|0\> ,
\ee
where $i = L$ or $T$, and it forms spin-1 representation.
% Here, ``level" means the sum of mode indices of the creation operators.
We obtain \be
	\alpha_{-1}^S|0\> = 0
	\label{L-1vac}
\ee
applying (\ref{aS}) with $n=-1$, which is the result of $L_{-1} |0\> = 0$.
This existence of 3 polarizations and absence of $S$ polarization agrees with the classical Nambu-Goto string.
Using $L'_{1} \alpha_{-1}^i|0\> = 0$, the invariant mass of the level-1 state is $p^2 = 2 - 2a$ because \be
	(L_0 - a)\alpha_{-1}^i|0\> = (-p^2/2 + 1 - a)\alpha_{-1}^i|0\> = 0.
\ee

We defined a physical state to be a state that fulfills $L_m|*\> = 0$ for any integer $m$.  The ground state is a physical state because $L_m|0\> = 0 \; (\forall m<0)$ are guaranteed by eq.(\ref{L-1vac}), eq.(\ref{L-2vac}) and so on.  No negative norm states will appear owing to these equations.
Other physical states are also guaranteed to be physical in the same way.
For the first example, it may be again proven using (\ref{aS}) that \be
	\alpha_1^S\alpha_{-1}^i|0\> = 0 .
	\label{aSa1}
\ee
Note that the commutator of $\alpha_1^S$ and $\alpha_{-1}^i$ is not assumed {\it a priori}.

In the second level, the physical states are \bea
	\alpha_{-1}^i \alpha_{-1}^i|0\> , 		\nn
	\alpha_{-2}^i|0\> , 			\nn
	\zeta_{ij}\alpha_{-1}^i \alpha_{-1}^j|0\> ,
	\label{l2phy}
\eea
where ${\rm tr}\zeta_{ij}= 0$ and each state has spin 0, spin 1, and spin 2 respectively.
The relations
\bea
	\alpha_{-2}^S|0\> &=&  {1\over 2} \alpha_{-1}^i \alpha_{-1}^i |0\> /p ,
	\label{L-2vac} \\
	\alpha_{-1}^S \alpha_{-1}^i|0\> &=& \alpha_{-2}^i|0\> /p,
	\label{L-1a1}
\eea
which comes from eq.(\ref{aS}), makes it possible to represent the other level-2 states: \ben
	\alpha_{-2}^S|0\> , 			\hspace{2em}
	\alpha_{-1}^S \alpha_{-1}^S|0\> , 			\hspace{2em}
	\alpha_{-1}^S \alpha_{-1}^i|0\> ,			\hspace{2em}
 	\alpha_{-1}^i \alpha_{-1}^S|0\>
\een
as linear combinations of the above physical states eqs.(\ref{l2phy}) or zero.  In this way, any product of $\alpha_n^i$'s and $\alpha_n^S$'s may be reduced to a product of only $\alpha_n^i$'s.

Let us obtain the invariant masses for the level-2 states.
For level-2 states, $L_0 = - p^2/2 + L'_0 -(L'_{-1}L'_1+L'_{-2}L'_2)/p^2 $ is sufficient owing to $L'_{-1}|0\> = 0$ and $L'_1 L'_1|\mbox{a level-2 state} \> = 0$.
The equation $(L_0-a)|*\> = 0$ gives $p^2/2 = 2 - a$ for the spin-2 state and
\be
	p^2 = 4 - 2a - {D-1\over p^2}
	\label{l2s0}
\ee
for the spin-0 state, whose solution is $p^2 \simeq 3$ for small $|a|$.  Here, $D$ is spacetime dimension.  For the spin-1 state in this level, we have \be
	p^2 = 4 - 2 a - 4/p^2 ,
\ee
and its solution is $p^2 \simeq 2$ for small $|a|$, which means this level-2 state have a mass close to that of the level-1 state, and nonzero $a$ will push the mass upward.
We do not take the other solutions of these quadratic equations.
Suppose that there are states with different masses.  Then the propagator for the particle is a sum of poles: \[
	\tilde\Delta(k^2) = \frac 1{k^2 + m_1^2} + \frac 1{k^2 + m_1^2} + \cdots .
\]
The zeroes of the effective action $\tilde\Delta(k^2)^{-1}$ corresponding to each pole have always positive slope in $k^2$ but not negative.  In analogy with this fact, we did not take the other solutions.
% Another reason is that the author tried other ordering rule than the normal ordering and found that the other solution with an energy lower than the half of the unperturbed energy $L_0'$ tend to be unstable.

Likewise, the level-3 physical states are \bea
	\alpha_{-3}^i|0\> ,
		\label{a3} 			\\
	\alpha_{-2}^i \alpha_{-1}^i|0\> ,
		\label{a2a1-0}			\\
	\epsilon_{kij}\alpha_{-2}^i \alpha_{-1}^j|0\> ,
		\label{a2a1-1}			\\
	\zeta_{ij}\alpha_{-2}^i \alpha_{-1}^j|0\> ,
		\label{a2a1-2}			\\
	\alpha_{-1}^i (\alpha_{-1}^j\alpha_{-1}^j)|0\> ,
		\label{a111-1} 			\\
	\zeta_{ijk}\alpha_{-1}^i \alpha_{-1}^j\alpha_{-1}^k|0\> ,
		\label{a111-3}
\eea
where each have spin 1, 0, 1, 2, 1, 3 respectively.  $\zeta_{ij}$ and $\zeta_{ijk}$ are totally symmetric traceless tensors.  The mass of the spin-3 state (\ref{a111-3}) won't be changed, i.e.\ $p^2 = 6 - 2a$ from $ L_0 - a $ constraint.
The states (\ref{a2a1-0}), (\ref{a2a1-1}) and (\ref{a2a1-2}) have the mass-shell conditions: \bea
	&& p^2/2 - 3 + a + (2(D-1)+4) / p^2 = 0,			\nn
	&& p^2/2 - 3 + a  = 0,			\nn
	&& p^2/2 - 3 + a + 4 / p^2 = 0
\eea
respectively.  The spin-0 state (\ref{a2a1-0}) will not appear since the condition for it does not have any solution.
The spin-1 states (\ref{a3}) and (\ref{a111-1}) get mixed by the operation of $L_0$ as \be
(L_0 - a) u
% = \Big\{ - p^2/2 + L'_0 -a - {1\over p^2} (L_{-1}L_1+ L_{-2}L_2) + {3\over p^4} (L_{-1}L_{-1}L_2 + L_{-2}L_1L_1) - {5\over 4p^6} L_{-1}L_{-1}L_1L_1) \Big\} u			\\
% \Big\{-{p^2\over 2} + 3 -a - {1\over p^2} A_3^T \Big( 1 - {3\over 2p^2}(P_3^T+ P_3) +{5\over 4p^4} P_3^T P_3 \Big) A_3 \Big\} u
= (- p^2/2 + 3 - a ) u
- {1\over p^2} \left( \begin{array}{cc}
	3(3 - 6/p^2 + 5/p^4)  &  (D+1)(1 - 3/p^2) 		\\
	3(1-3/p^2)/2  &  (D+1)/2
	\end{array} \right) u ,
	\label{l3s1}
\ee
where \[
	u = \left( \begin{array}{c}
		\alpha_{-3}^i|0\> 		\\
		\alpha_{-1}^i (\alpha_{-1}^j\alpha_{-1}^j)|0\>
	\end{array} \right) .
\]
\delete{
where \ben
	A_3 = \left(\begin{array}{cc}
		\sqrt{3} & \sqrt{(c+2)/2} \\
		\sqrt{6} & 0
	\end{array} \right),			\\
	P_3 = \left( \begin{array}{ccc}
		0 & \sqrt{c/2} 			\\
		0 & 0
	\end{array} \right) .
\een
}
The $L_0$ constraint will be fulfilled when it has zero eigenvalue, and then $p^2$ will be determined.
Numerical solution may be obtained for this.
For example, $p^2 \sim 3.1$ and $p^2 \sim 5.4$ is the solutions when $a=-0.1$.

We will now show the calculation required for obtaining the level-4 spin-0 mass as an example of level-4.
% We will see an example of increasing complication into higher orders.

At level 4, there are three spin-0 states: \be
v = \left(\begin{array}{c}
	 \alpha_{-3}^i \alpha_{-1}^i|0\> /{\sqrt{3c}}			\\
	 \alpha_{-2}^i \alpha_{-2}^i|0\> /{\sqrt{2c}}			\\
	 (\alpha_{-1}^i \alpha_{-1}^i)^2|0\> /{\sqrt{2c(c+2)}}
	\end{array} \right)
\ee
where $c = D-1$.

Then these states should be linearly recombined to give a zero eigenvalue for $L_0$, and this condition determines the shell $p^2$.
The states in $v$ are mixed by $L_0$ (\ref{L0}) as \be
	(L_0 - a) v = \Big\{-{p^2\over 2} + 4 -a
		- {1\over p^2} A^T \Big( 1 - {3\over 2p^2}(P^T+ P) +{5\over 4p^4} P^T P \Big) A \Big\} v ,
	\label{l4s0}
\ee
% = \Big\{ - p^2/2 + L'_0 -a - {1\over p^2} (L_{-1}L_1+ L_{-2}L_2 + L_{-4}L_4) + {3\over p^4} (L_{-1}L_{-1}L_2 + L_{-2}L_1L_1 + L_{-2}L_{-2}L_4 + L_{-4}L_2L_2) - {5\over 4p^6} L_{-1}L_{-1}L_1L_1 + L_{-2}L_{-2}L_2L_2) \Big\} u			\\%  not checked yet
% end comment
where \ben
	A &=& 	\left(\begin{array}{ccc}
		0 & \sqrt{6} & \sqrt{c+2} \\
		2 & \sqrt{6} & 0 		\\
		\sqrt{2c} & \sqrt{3c} & 0
	\end{array} \right),			\\
	P&=&  \left( \begin{array}{ccc}
		0 & 2 & 0 			\\
		0 & 0 & 0 			\\
		\sqrt{c/2} & 0 & 0
	\end{array} \right) .
\een
This equation (\ref{l4s0}) was numerically solved.
For example when $a = -0.1$, one of the solutions was $p^2 \sim 5.6$ and the zero-eigenvalue eigenvector was close to $(-0.6, 1.0, 1.6)\cdot v$, which is the third spin-0 state.
The other solution has too large energy to plot in Fig.\ref{fig:meson} ($p^2 \sim 8.0$).
% For $a= -0.3$, which corresponds to $\eta'$ meson, the solution was $p^2 \sim 6.0$.

\section{ Application as a phenomenological model of hadrons }

Let us consider our theory as a phenomenological model of mesons in this subsection, though this theory may also be used as a theory of fundamental strings.
% This model is appropriate for massless quarks at the end of the strings.

\begin{figure}[tbp]
 \begin{center}
  \includegraphics[width=140mm]{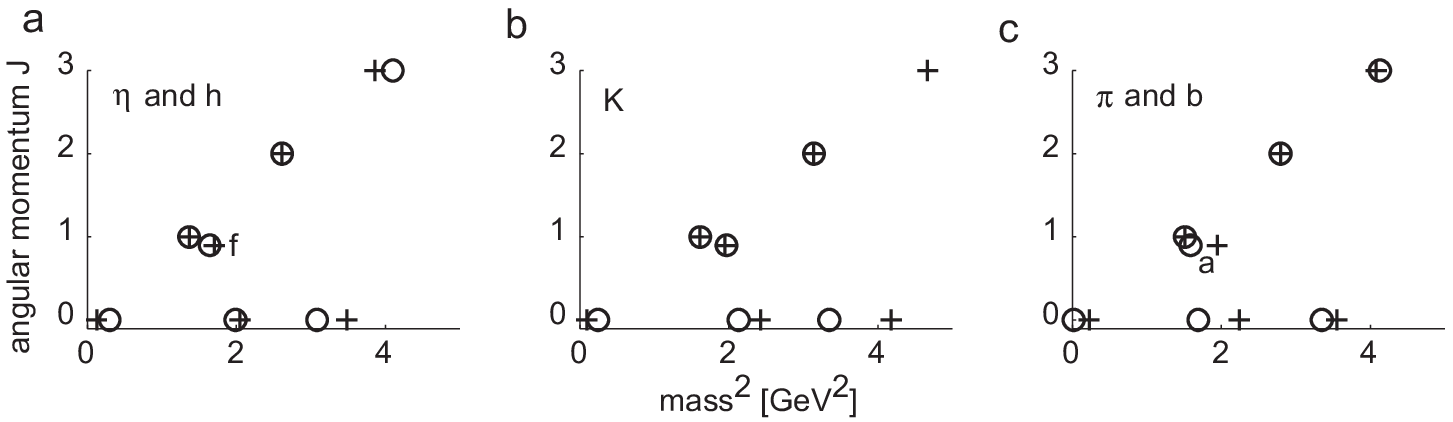}
 \end{center}
 \caption{
The experimental masses (circles) of light mesons and our theoretical predictions (crosses) are plotted for a) $\eta$ and $h$ mesons ($I=0$), b) Kaons ($I=1/2$), and c) pi-on family ($I=1$).
The meson masses are taken from ``review of particle physics''\cite{PDG08}.
}
 \label{fig:meson}
\end{figure}

Our theoretical result and experimentally known spectra were plotted in fig.\ref{fig:meson}.
The string tension and the variable $a$ are determined to fit to the lowest energy states of spin-1 and spin-2, for example, to $K_1(1270)$ and $K_2(1430)$.
These states were chosen since our stringy picture should most deviate from QCD at the ground state, and our theory gives the simplest form at the least mass state for each angular momentum.
% and the correction from Coulomb force should be larger in S orbit states and largest at the ground state.

The plots of the calculated spectra and the observed spectra in fig.\ref{fig:meson} show striking agreement.
We see especially good agreement in fig.\ref{fig:meson}a for $\eta$ and $h$ mesons that mainly decay into nonstrange mesons.
% For pi-on, the theoretical and experimental masses are plotted in fig.\ref{fig:meson}c.

The most remarkable agreement is that we have heavier particles for each of spin value.\footnote{
In the conventional quantization methods of the strings, such as light-cone gauge or BRST formalism, there was only lightest particle for each angular momentum. % GSW p114
% lacks the excited states for each of angular momentum L
% leaving only the Regge trajectory.
}

A notable feature of eq.(\ref{l2s0}), which determines the mass of the spin-0 level-2 particle, is that the mass gets smaller than the spin-2 level-2 particle.  The mass looks to agree the masses of the real mesons and especially well with that of $\eta(1405)$ and $K(1460)$.
This effect is reminiscent of that the energy of $L= 0$ state of a hydrogen atom gets lower by relativistic correction.

Another notable feature of our prediction is that the level-2 $\alpha_{-2}^i|0\> $ state has a mass close to level-1 $\alpha_{-1}^i|0\> $ state.
In this 2nd vibration mode, the string is bent at the middle and the ends do not move relative to each other.
Therefore this state have $J^{PC} = 1^{++}$, whose charge conjugation quantum number $C$ is opposite to that of ordinarily supposed $L=1$ quark bound states, since the relative position of the quarks at the ends takes the S orbit.
In $I=0$ family (Fig.\ref{fig:meson}a), the particle that match the quantum number $1^{++}$ is $f_1$ and not $h_1$, and  $f_1(1205)$ matches the calculated mass well.
For Kaons, since they don't have $C$ quantum number, $K_1(1400)$ may be identified.
In pi-on family $(I=1)$, the $a_1$ mesons is appropriate to identify to this state.  We tentatively identified $a_1(1260)$.
% a_1(1640) another less likely candidate

We see some discrepancy in the masses of spin-0 states in the Kaon spectrum (fig.\ref{fig:meson}a).  This discrepancy may be the result of the strange quark at one of the ends, since the $\eta$ and $h$ mesons that mainly decay into strange mesons had shown further larger deviation (not plotted).

% In the level-3 spin-0 state, no solution was there, which is also a good news.  This state have $0^{--}$ quantum number, which is not experimentally found for nonstranged mesons ($\eta$ and $\pi$).  % though we have $\phi(1680)$.

The following states that are not plotted in the figure also have shown some agreement.
The third $L=1$ state, which is the lower energy solution of eq.(\ref{l3s1}), seems to correspond to $h_1(1595)$ state but does not have counterpart among $K$ and $b$ mesons.
The second $L=2$ state (\ref{a2a1-2}) gives 1796MeV for the Kaon and seems to correspond to $K_2(1820)$, which is 1 percent precision, though isospin-zero and isospin-1 mesons do not have counterpart, which may mean that they can be experimentally found in the future.
The other solutions, i.e. the higher energy solution of eq.(\ref{l3s1}) ($J^{PC} = 1^{+-}$) and the state (\ref{a2a1-2}) (abnormal $J^{PC} = 1^{--}$), does not seem to have corresponding particles experimentally so far.
This absence seems to be because these energy regions are not well explored.

We did not tried to identify the states in that the pair of quark spins form a spin $S=1$ representation, since they are three times numerous theoretically and experimentally making it difficult to identify the correspondence, and we wanted to stay the safer side.

\delete{
Fig.1d plots the mesons that mainly decay into $K\bar K$.  We also see fairly good agreement.
The mass of $\eta_0$ should be pushed down and $\eta_0'$ should be pushed up, because transitions can occur between these mesons and the closed string states, and each of the two states pushes energy of the other states away in general.
This is because emergence of a closed string is inevitable since the ends of a open string can merge by a pair annihilation of the quarks for these isospin $I=0$ states.  We considered $f_0(600)$ meson as the ground state of a closed string.
% f_2(2340) seems to be a good candidate as the first excited state of a closed string, which should have twice as heavy as the first excited states of the open strings, but it may be premature to identify.
}

The experimental and theoretical spectra have shown good agreement.  Giving correction about strange quark mass at the ends may further improve the agreement.

\section{ Discussion }

In this paper, we presented a method to quantize the string action.
This string may be considered to be either a fundamental string or a phenomenological string of hadrons, i.e.\ of the QCD.

% The following extensions will immediately be possible.

A trivial extension is applying it to closed strings.  There is no obstacle in that.
Emergence of a closed string is inevitable since the ends of a open string can merge by a pair annihilation of the quarks for the isospin $I=0$ states.  The $f_0(600)$ meson may be the ground state of the closed string.
% f_2(2340) seems to be a good candidate as the first excited state of a closed string, which should have twice as heavy as the first excited states of the open strings, but it may be premature to identify.

Further, it would not be difficult to quantize a string with a heavy particle at one of the ends.  Then the spectra of the heavy-quark mesons won't be difficult to obtain.
It would also not be difficult to quantize Nambu-Goto action for three-armed (trident-like) string, and the baryon spectra can be obtained.

Supersymmetrized strings are also possible in our formalism though we don't have to reduce the dimension to 10.  It may be one of ways to introduce a massless spin-2 particle, i.e.\ graviton.
It seems also possible to quantize extended objects in any dimension such as membranes.

Our formula is showing agreements and similarities with the effective string theory.  The effective string theory considers long string, in which problem of conformal anomaly can be evaded, as an effective theory of QCD strings.  In our theory, a long string may be considered to be a highly excited state in a direction, say $(\alpha_1^x)^{R^2}|0\>$, where $R$ is the mean length.  The mass is roughly $ p^2 \sim R^2$.  Let us suppose that we have further $n$ excitations that are perpendicular.
Our formula (\ref{L0}) and $L_0 - a = 0$ gives a formula for the squared mass $p^2$.  It may be rewritten as \be
	p^2 = R^2 + n - a - {1\over p^2}  \sum_m L_{-m} L_m + O(1/R^4) .
\ee
Putting this formula itself into $p^2$, taking a square root of the both sides and expanding, we get \be
	p = R  + {n-a \over 2 R } - {1\over 2 R^3} (\sum_m L_{-m} L_m + (n-a)^2/4)+ \cdots  .
\ee
% The second term is the familiar L\"uscher's correction.
This agrees with L\"uscher and Weisz's result\cite{Lush04}, which is an effective string theory in static gauge, on that the second correction is $1/R^3$ correction and its factor $ L_{-n} L_n$ is related to the second term of the expansion of Nambu-Goto action, i.e.\ $(\partial_a h \partial_a h)^2$.  Further it agrees that a split of energy by the angular momentum starts at this order.
Thus similarity is clear; however, the argument above still does not completely reproduce fixed boundary condition of open strings or periodic boundary condition into elongated direction, which are typically used in the effective string theory.  Therefore further elaboration is needed to see the agreement in detail.

Our method of quantization by Euler-Lagrange equations of motion is a well-defined and consistent way, then it may also serve as a mathematical definition of quantum field theories.
%
% Applying this quantization method to general relativity will give a consistent theory of quantum gravity.  The trouble with gravity was that we had trouble in handling a theory with diffeomorphism invariance, which is solved in our string theory, and the same method may be applied to the gravity.

A problem that should be worked out in the future is construction of the interaction vertices and showing its S-duality.
%
% In this theory, the four-point amplitude will have S-duality though the spectrum and amplitude looks too complicated.

Here we succeeded in quantizing the string, and the light-meson spectra were accounted.  This method will be further useful in elucidating physics of gauge symmetric systems.

\appendix

\section*{Appendix A: A reconsideration on quantization }
% A new quantization method
% What is Quantization ?

Let us slightly shift our understanding or view about what quantization is.  This new method is equivalent to canonical quantization in ordinary systems but gives slightly different results in constrained systems.

In the canonical quantization, we first determined the representation by setting the commutation relations and then derived the equations of motion in Heisenberg picture.
We easily see below that the inverse is possible and equivalent, i.e. we may first give the equations of motion and can derive the commutation relations.

Now let us define quantization as:
\begin{definition}
% \label{quant0}
Quantization is finding a nontrivial representation of a non-commutative algebra % of configuration space variables
in which all the equations of motion hold as operator equations.
\end{definition}

Then, for a non-constrained system, its quantization was complete if you can find such a space in that all the states obey the equations of motion \[
	\Big[ {d\over dt} - iH, \phi \Big]|*\> = 0
\]
in Heisenberg picture.
% or equivalently d/dt|*> = H|*\> in Schr\odinger picture.

Let us see that this modified method of quantization is equivalent to the conventional one, for example, in the free scalar field theory \[
	L = \int d^3x {1\over 2} \Big\{ (\d_0 \phi )^2 - (\d_i \phi )^2 -  m^2 \phi^2(\v x) \Big\}
\]
as follows.  The equation of motion for this Lagrangian is \be
	(\d_0^2 - \d_i^2) \phi(x)  = - m^2 \phi (x) .
	\label{phi-3}
\ee
According to our definition, quantization is requiring the equation of motion (\ref{phi-3}) to be an operator equation.  For the operator equations, there should exist a solution $\phi (t,\v x)$, which is written in terms of $t = 0$ variables $\phi (0,\v x)$ and $\dot \phi (0,\v x)$.  Now let us consider a commutator \[
	[\phi (t,\v x), \phi (0,\v x')] \equiv  i\Delta (x, x'),
\]
which, by definition, obeys \be
	(\d_0^2 - \d_i^2 + m^2) \Delta (x, x') = 0.
	\label{1}
\ee
We would like to find minimally nontrivial algebra consistent with our definition except for the obvious and trivial solution $\Delta (x, x') = 0$, which corresponds to the classical field theory.
% Let us further assume that $\Delta (x, x')$ is c-number valued for commutators of fundamental variables that form phase space.  This assumption is looser than that of canonical quantization.
%
Now we have \[
	\Delta (x, x') = \Delta (x-x')
\]
because of translation invariance, and from eq.(\ref{1}) we have \[
	\tilde\Delta (k) = c \; {\rm sign}(k_0) \delta( k_0^2 - \v k^2 - m^2 ),
\]
where ${\rm sign}(k_0)$ was required since $\Delta (x)$ must be an odd function, and
$ c $ is an undetermined factor of dimension zero, which can be a operator.  Let us suppose that this dimension zero operator, which is translation invariant, is a c-number.
Differentiating the Fourier invert: \[
	\Delta (x) = c \int {d^4 k\over (2\pi)^3 i} {\rm sign}(k_0) \delta( k_0^2 - \v k^2 - m^2 ) e^{-i k x}
\]
by time and setting $t = t'$, we obtain \be
	[\dot \phi (t,\v x), \phi (t,\v x')] = c \;\delta^3(x - x').
	\label{phicomm}
\ee
The undetermined factor $ c $ can be absorbed into a rescale of the field.  Therefore this undeterminedness does not matter.
Please note that $ c $ could have been dependent on $ k$, which possibility was turned down by Lorentz invariance of the system.  Therefore requiring adequate symmetry is a part of the definition.
% Here the locality of the commutators is deduced only from Poincare invariance without explicitly requiring it.
%
The representation of this system follows this commutation relation (\ref{phicomm}).  First, define annihilation operators and creation operators as linear combinations of $\phi (\v x)$ and $\pi (\v x)$, then define the vacuum as \be
	a_k|0\> = 0
\ee
for all $\v k$.  Then all the other states are defined by operating creation operators on this vacuum.  This forms a representation for the algebra defined by the equation of motion.
%
% In our string quantization in contrast, the algebra will rather be determined in the course of determining the representation.

In short, we have usually started with the commutation relation, e.g.\ (\ref{phicomm}), and derived equation of motion in Heisenberg picture.  Now we see that, starting from the equation of motion, we can derive the commutation relation inversely.

A constrained system is a system that does not have as many time derivatives $\dot x_i$ as the number of the configuration variables $x_i$ in the Lagrangian, and then have difficulty in moving to canonical formalism.
The simplest example of constrained systems is $L = x^2/2$, in which the canonical momentum $p = \d L/\d \dot x = 0$.  Such a condition for canonical variables as this $p = 0$ is called a ``constraint''.

Now let us define quantization, allowing for constrained systems, as:
\begin{definition}
\label{quant}
Quantization is finding a nontrivial representation of a non-commutative algebra % of configuration space variables
in which all the equations of motion and the constraints hold as operator equations.
\end{definition}
Here the representation should conform to the naturally expected symmetries such as Poincare invariance, gauge symmetry.
Treating the equations of motion and constraints on an equal footing is natural because they come out on an equal footing in the Euler-Lagrange equations.

We already reported \cite{nonfix} that this method successfully quantized non-Abelian gauge fields, which are constrained systems, without any additional trick of gauge fixing and BRST formalism.

Further, the treatment of first-class systems becomes simpler in our formalism.  There were complication coming from indeterminacy, which is the defining feature of the first-class constrained systems, and we had to handle it with gauge fixing and Faddeev-Popov determinant.
Now we don't have to do anything with the indeterminacy, which is same as that we did nothing with the indeterminacy of gauge in classical electrodynamics and all the solution that conformed to the equations of motion were equivalent physically.

Not only our method can treat constrained systems and nonconstrained systems in the same way, our method can also treat constrained systems of the first and the second class in an equal way in contrast to Dirac's prescription\cite{Dir64}, which is considered to be the standard way to do with the constrained systems.
The second-class systems had to be treated differently because the second-class constraints are incompatible with the canonical commutation relation.

Let us see how the trouble with the second-class systems was solved in the simplest example.
For a nonconstrained system $L = {\dot x}^2/2 - x^2/2$, our quantization method gives operator equations: \ben
	&& \dot p = -x ,			\\
	&& \dot x = p ,
\een
and their solution exists as $x(t) =  x(0) \cos t + p(0) \sin t $.  Let us consider a commutator \[
	[x(t), x(0)] \equiv  c(t, 0).
\]
From definition, we have \[
	\d_t^2 c(t, t')  = 0,
\]
and then \be
	[\dot x(t), x(0)] \equiv  {d\over dt} c(t) = {\rm const}.
\ee
Thus this system can have a nonvanishing commutation relation.
In contrast, considering the simplest second-class system $L = {1\over 2}x^2$, we have \bea
	x|*\> = 0,			\\
	p|*\> = 0,
\eea
as equations of motion, and they naturally leads to \be
	[x, p] = 0
\ee
without any additional assumption.\footnote{
The Dirac bracket was invented to make the canonical variables to be $[x, p] = 0$.
The trouble of requiring different method for the second-class constrained systems was because canonical commutation relation was introduced first and then the constraints were imposed in the Dirac's method.  Our method has imposed the constraints first, then commutation relations are determined consequently.
}
% When we have multiple variables, our method can determine the subspace that should be reduced with $[x, p] = 0$.

\begin{figure}[tbp]
 \begin{center}
  \includegraphics[width=140mm]{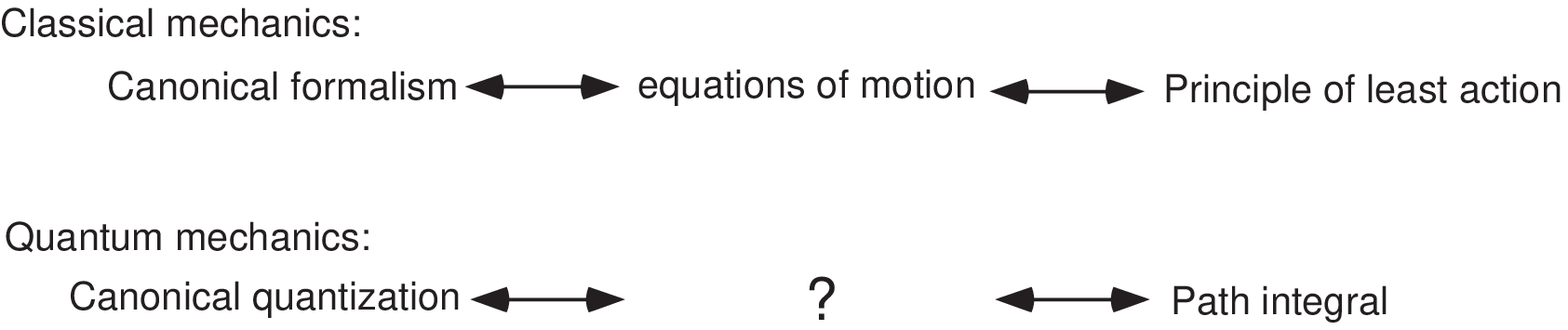}
 \end{center}
 \caption{What fits at the position of the question mark?}
 \label{fig:eqmot}
\end{figure}

Our formalism fit into the place of equation of motion in classical mechanics (Figure.\ref{fig:eqmot}). % 10.3.17
In classical mechanics, the equations of motion were first there, and Lagrange's formalism and Hamilton's formalism (canonical formalism) were invented later.
If one wanted to move from Lagrange's formalism to canonical formalism, we once deduce Euler-Lagrange's equations of motion and then canonical formalism was derived.  We do the same also in the opposite movement.
In quantum mechanics, we knew canonical quantization and path integral, which corresponds to Lagrange's formalism.  However, something that corresponds to the equation of motion was not known.
Our formalism fits in this place.
These three formalisms in quantum mechanics are equivalent to each other in ordinary (nonconstrained) systems, but acts differently to systems with gauge symmetry.
% Canonical formalism gives constraints, path integral diverges, our formalism acts normally.
Our method has advantage in this respect.

\section*{Appendix B: Lorentz Covariance }

In the conventional lightcone-gauge quantization, Lorentz invariance was broken, and $D = 26$ was required to restore the algebra of Lorentz generators.
Therefore some of the reader may think that our theory should also have the same problem and be breaking Lorentz invariance somewhere.
It may be better to answer to this argument, even though our theory is manifestly Lorentz invariant because every step in the derivation is covariant.

Here we explicitly show that such linear operators exist that produce Lorentz transformations for our system: \be
	[M_{\mu \nu }, M_{\rho \sigma}]  =  i\eta _{\mu \rho } M_{\nu \sigma} - i\eta _{\mu \sigma} M_{\nu \rho} - i\eta _{\nu \rho} M_{\mu \sigma}+ i\eta _{\nu \sigma} M_{\mu \rho} .
	\label{LzAlg}
\ee

We have three polarizations $+, -,$ and $L$.  The L-polarization vector is \be
	\e^L = (|\v p|, p_0 \v e^{\rm p}) / p,
\ee
where $\v e^{\rm p} = \v p/|\v p|$, and $p = \sqrt{p_0^2-\v p^2}$.  It is dependent on $\v p$ and we will write it as $\e^L(p)$ when we want to stress this dependence.  The transverse polarizations are dependent on $\v p$ as well.   They have only spatial components, and let us denote them by $	\v e^2, \v e^3, ..., \v e^{D-1}  $
supposing that they are linearly polarized and orthogonal to each other.   In three dimensions, they are related to $\v e^+ = (\v e^2 + i\v e^3)/\sqrt 2$ and $\v e^- = (\v e^2 - i\v e^3)/\sqrt 2$.  There is arbitrariness how to choose the set $\{\v e^l \}$ on the sphere $|\v p| = {\rm const}$.

Let us first consider an infinitesimal Lorentz transformation in the classical (and quantum) theory.  The total momentum will be transformed by \be
	(p^0, \v p) \rightarrow  (p^0 + \v \beta \!\cdot\!\v p, \v p + \v \beta p^0)
\ee
with a infinitesimal boost $\v \beta $.  The vector $X^\mu $ should also be transformed according to the same Lorentz boost \bea
	&& X'^0 = X^0 + \beta^i X^i ,			\nn
	&& X'^i = X^i + \beta^i X^0.
	\label{LV}
\eea
Therefore for a longitudinal polarization $\alpha^L_n$, its polarization vector $\e^{L\mu}$ is supposed to transform as \be
	{1\over  p} (|\v p|,  p^0 \v e^{\rm p})
	\rightarrow   {1\over  p} (|\v p| + \v \beta \!\cdot\!\v e^{\rm p} p^0, p^0 \v e^{\rm p} + \v \beta |\v p|) \equiv {\e^L}'
	\label{boostedPola}
\ee
to the first order in $\v\beta $.  After this change of momentum $\v p$, the $L$ polarization direction changes to $
	(|\v p + \v \beta p^0|, \v e_{p'}  (p^0 + \v \beta \!\cdot\!\v p)) /p , $
where $\v e_{p'} = \v p'/|\v p'| = (\v p + \v \beta p^0)/|\v p + \v \beta p^0|$.  Then the polarization (\ref{boostedPola}) should again be decomposed into $L$ and $T$ polarizations: \be
	{\e^L}' = \e^L(p') - {p \over |\v p|}\sum_l (\v \beta \cdot \v e^l ) \e^l(p')  .
\ee
Let us next consider a transverse polarization $\e = (0, \v e)$.  It is transformed by the Lorentz boost as $
	(0, \v e) \rightarrow  (\v \beta \cdot \v e, \v e)		$
and this polarization vector should be decomposed into new $L$ and transverse polarizations: \be
	{\e^l}' = \e^l(\v p') + (\v \beta \cdot \v e^l) {p\over |\v p|}\e^L(\v p')
		+ \sum_m \v e^l\cdot {\d \v e^m\over \d p_j} \beta _j p_0 \e^m(\v p').
\ee
The transformation for creation and annihilation operators may be read by putting \ben
	X^\mu (\v p) &=&  i\sum_{P: {\rm polarizations}} \e^{P\mu}(\v p) (\alpha^P - \alpha^{P\dagger}) ,		\\
	X'^\mu (\v p') &=& i\sum_{P: {\rm polarizations}}  \e^{P\mu}(\v p') (\alpha '^P - \alpha '^{P\dagger})
\een
into eq.(\ref{LV}) to obtain \bea
	\alpha'^L_n &=& \alpha^L_n + {p\over |\v p|} \sum_l \v \beta \cdot \v e^l \alpha^l_n  ,		\nn
	\alpha'^l_n  &=& \alpha^l_n - {p\over |\v p|} \v \beta \cdot \v e^l \alpha^L_n
			- \sum_m \v e^l\cdot {\d \v e^m\over \d p_i} \beta _i p_0 \alpha^m_n .
	\label{L-a}
\eea
The discussion up to here holds for both classical and quantum strings.

The transformation (\ref{L-a}) is realized by the following generator. \be
%	K_\beta  = \beta_i M_{i0}
	M_{i0} = (x_i p_0 - x_0 p_i)  - i \sum_n \Big\{
			 {p\over |\v p|} e^l_i (\alpha^{l\dagger}_n \alpha^L_n - \alpha^{L\dagger}_n \alpha^l_n)
			+\v e^m\cdot {\d \v e^l\over \d p_i} p_0 \alpha^{l\dagger}_n \alpha^m_n
		\Big\},
\ee
where the indices $l, r$ are spatial indices excluding $L$ and run from 2 to $D-1$, and the Cartesian spatial indices $i,j$ can take from 1 to $D-1$.  The index $n$ and $\sum_n$ will be abbreviated from now on.  Repeated indices are supposed to be summed.  $ X_0^\mu$ and $ \Pi_0^\mu$ are here written as $ x_\mu$ and $ p_\mu$, and obey $[x_\mu, p_\nu] = i \eta_{\mu\nu}$.
With the same argument as above, the rotation generator should be \be
	M_{ij} = (x_i p_j - x_j p_i) - i \Big( e^l_i e^m_j +\v e^m\cdot {\d \v e^l\over \d p_i} p_j - (i \leftrightarrow j)\Big) \alpha^{l\dagger} \alpha^m ,
\ee
which is consistent with that the $L$ polarization will not be changed because it is parallel to $\v p$, and that the transverse components will be rotated.  The $\v e^m\cdot {\d \v e^l\over \d p_i} p_j$ term comes from that the direction of $\v e^l$ depends on $\v p$ and components in this direction should be corrected when $\v p$ is changed.

One can confirm that these operators reproduce the correct Lorentz transformation (\ref{L-a}) by boost as \bea
	&& [M_{i0}, M_{j0}] = i (x_i p_j -x_j p_i)
	+ {p^2\over |\v p|^2} e^l_i \alpha^{l\dagger} \v e^m_j \alpha^m			\nn
	&& - \Big( - {p\over |\v p|} e^r_i \alpha^{L\dagger} + \v e^r\cdot {\d \v e^l\over \d p_i} p_0 \alpha^{l\dagger} \Big) \Big( {p\over |\v p|}  e^r_j \alpha^L + \v e^m\cdot {\d \v e^r\over \d p_j} p_0 \alpha^m \Big) 		\\
	&& + p_0 {\d \over \d p_i} \Big( {p\over |\v p|} e^l_j  (\alpha^{l\dagger}\alpha^L - \alpha^{L\dagger}\alpha^l ) +\v e^m\cdot {\d \v e^l\over \d p_j} p_0 \alpha^{l\dagger} \alpha^m \Big)
	- (i \leftrightarrow j)			\nonumber
\eea
using the commutation relation (\ref{comm}), with $r = 2, ... , D-1$.  Here $e^l_i$ denotes the $i$-th component of vector $\v e^l$.  The equation \be
	[M_{i0}, M_{j0}] = - i M_{ij}
\ee
may be confirmed with the use of \bea
	\v e^l\cdot {\d \v e^m\over \d p_i} &=& -\v e^m\cdot {\d \v e^l\over \d p_i},			\\
	\v p\cdot {\d \v e^m\over \d p_i} &=& - e^m_i ,
	\label{r1}
\eea
which are derivatives of $\v e^l\cdot \v e^m = 0$ and $\v p\cdot \v e^m = 0$. Further, we have \be
	\sum_l e^l_i e^l_j = \delta_{ij} - e^{\rm p}_i e^{\rm p}_j .
\ee
Then it is easy to check on \bea
	[M_{ij}, M_{k0}] &=&  i\delta_{ik} M_{j0} - i\delta_{jk} M_{i0} ,			\\ \,
	[M_{ij}, M_{kh}]  &=& i\delta_{ik} M_{jh} - i\delta_{ih} M_{jk} - i\delta_{jk} M_{ih}+ i\delta_{jh} M_{ik} .
\eea
with those relations, which completes the proof of eq.(\ref{LzAlg}).

% It is same as representation of a single particle with polarization vector represented in helicity components.

\end{document}